\documentclass[aps,twocolumn,pra,groupedaddress,superscriptaddress,amsmath,amssymb,amsthm]{revtex4-1}
\usepackage{subfigure,hyperref,bbm,times}
\usepackage[T1]{fontenc}
\usepackage{braket}
\usepackage{amsthm}
\usepackage{epsfig}
\usepackage{color}
\usepackage{graphicx}% Include figure files
\usepackage{dcolumn}% Align table columns on decimal point
\usepackage{bm}% bold math
\usepackage{cancel}
\usepackage{bm}

\providecommand{\openone}{\leavevmode\hbox{\small1\kern-3.8pt\normalsize1}}

\begin{document}

\title{Dealing with indistinguishable particles and their entanglement}

\author{Giuseppe Compagno} 
\email{giuseppe.compagno@unipa.it}
\affiliation{Dipartimento di Fisica e Chimica, Universit\`a di Palermo, via Archirafi 36, 90123 Palermo, Italy}
\author{Alessia Castellini}
\affiliation{Dipartimento di Fisica e Chimica, Universit\`a di Palermo, via Archirafi 36, 90123 Palermo, Italy}
\affiliation{INFN Sezione di Catania, Catania, Italy}
\author{Rosario Lo Franco}
\email{rosario.lofranco@unipa.it}
\affiliation{Dipartimento di Fisica e Chimica, Universit\`a di Palermo, via Archirafi 36, 90123 Palermo, Italy}
\affiliation{Dipartimento di Energia, Ingegneria dell'Informazione e Modelli Matematici, Universit\`{a} di Palermo, Viale delle Scienze, Edificio 9, 90128 Palermo, Italy}

\date{\today }% It is always \today, today,

%\pacs{03.65.Ta, 03.67.-a, 03.65.Ud}% PACS, the Physics and Astronomy

\begin{abstract}
Here we discuss a particle-based approach to deal with systems of many identical quantum objects (particles) which never employs labels to mark them. We show that it avoids both methodological problems and drawbacks in the study of quantum correlations associated to the standard quantum mechanical treatment of identical particles. The core of this approach is represented by the multiparticle probability amplitude whose structure in terms of single-particle amplitudes we here derive by first principles. To characterise entanglement among the identical particles, this new method utilises the same notions, such as partial trace, adopted for nonidentical ones. We highlight the connection between our approach and second quantization. We also define spin-exchanged multipartite states (SPES) which contain a generalisation of W states to identical particles. 
We prove that their spatial overlap plays a role on the distributed entanglement within multipartite systems and is responsible for the appearance of nonlocal quantum correlations.
\end{abstract}

\maketitle

\section{Introduction}

Identical quantum objects (e.g., qubits, atoms, quantum dots, photons, electrons, quasiparticles), tipically are the basic "particles" forming the building blocks of quantum-enhanced devices \cite{laddReview,bloch2008many,anderlini2007controlled, exptenphoton,fivequbitexp,RevModPhys.81.1051,benatti2013PRA,crespi2015,Martins2016PRL, martinisfermions,QEMreview}. Characterising the quantum properties of composite systems of identical particles is hence important for both fundamental and technological aspects. In quantum mechanics, due to their indistinguishability, identical particles are not individually addressable and require specific treatments which differ from those used for nonidentical (distinguishable) particles.  

In the standard quantum mechanical approach (SA) to deal with identical particles the first step is to assume they are not, marking them with \textit{unobservable} labels \cite{peresbook,cohen2005quantum}. This is the only place where non observable quantities occurs, being quantum states defined by complete sets of commuting observables. SA requires that the system is described by states whose structure is constrained to be symmetric (bosons) or antisymmetric (fermions) with respect to the labels \cite{feynman,cohen2005quantum,goyal2014}. So an intrinsic entanglement is present, even for independently prepared separate particles, which is attributable to the particle fundamental indistinguishability. This has given rise to different viewpoints about physical meaning and assessment of this part of entanglement \cite{balachandran2013PRL,plenio2014PRL,ghirardi2004PRA, schliemann2001quantum,eckert2002AnnPhys,wiseman2003PRL, dowlingPRA,vaccaroIJQI,jonesPRA,buscemi2007PRA,vogel2015PRA, giulianoEPJD, benatti2012AnnPhys,sasaki2011PRA,benatti2012PRA, benattiOSID2017,franco2016quantum,sciara2017universality}.  
For some, the entanglement due to indistinguishability is considered present but unusable in a state of independently prepared distant particles \cite{peresbook}, for others it is seen as a merely formal artifact even when the particles are brought to overlap \cite{tichy2011essential,ghirardi2004PRA}. Establishing the physical nature of identical particle entanglement is therefore crucial to  identify its role as a resource for quantum information and communication processing \cite{horodecki2009quantum, giovannetti2004Science,riedel2010Nature,benatti2014NJP, cramer2013NatComm,marzolino2015,marzolino2016,  dowlingPRA,QEMreview,plenio2014PRL,LFRoperationalframe}.

Recently a different approach has been introduced \cite{franco2016quantum} which describes the quantum states of identical particles without introducing unobservable labels. This non-standard approach (NSA), so far limited to two identical particles, exhibits peculiar advantages both from conceptual and practical viewpoints, linked in particular to the treatment of their quantum correlations. In fact, it from the beginning avoids the existence of the entanglement due to unobservable labels and it allows the quantification of two-particle entanglement by means of the same notions employed for distinguishable particles, such as the partial trace. This NSA has permitted to show: the existence of the Schmidt decomposition for identical particles \cite{sciara2017universality} (showing that it is universally valid both for identical and not identical particles); a new efficient generation scheme of multipartite W entangled states \cite{BLFC2017} and the definition of an operational framework to directly exploit entanglement due to indistinguishability for quantum information protocols \cite{LFRoperationalframe}. Here we reconsider the NSA from a fundamental perspective and generalise it to a system of many identical particles.

The paper is organized as follows. In Section~\ref{Sec:SA} we briefly present, as an example, some immediate problematic implications of the use of the quantum SA for the characterization of the entanglement between identical particles. In Section~\ref{Sec:NSA} we give the formalism and the tools of the non-standard approach (NSA) for a system of $N$ identical particles in a pure state. We obtain the many-particle probability amplitudes from first principles. In Section~\ref{Sec:GenStates} we generalise states of nonidentical particles to system of identical ones. In Section~\ref{Sec:App}  we evidence the role of the spatial overlap in the entanglement evaluation for three identical qubits, and the Bell inequality violation for independently prepared identical particles within an operational framework. We finally summarize our conclusions in Section~\ref{Sec:Conc}.

\section{Problematics of the standard approach (SA) to identical particles}
\label{Sec:SA}

\begin{figure*}[!t]
\centering\includegraphics[width=0.6\textwidth]{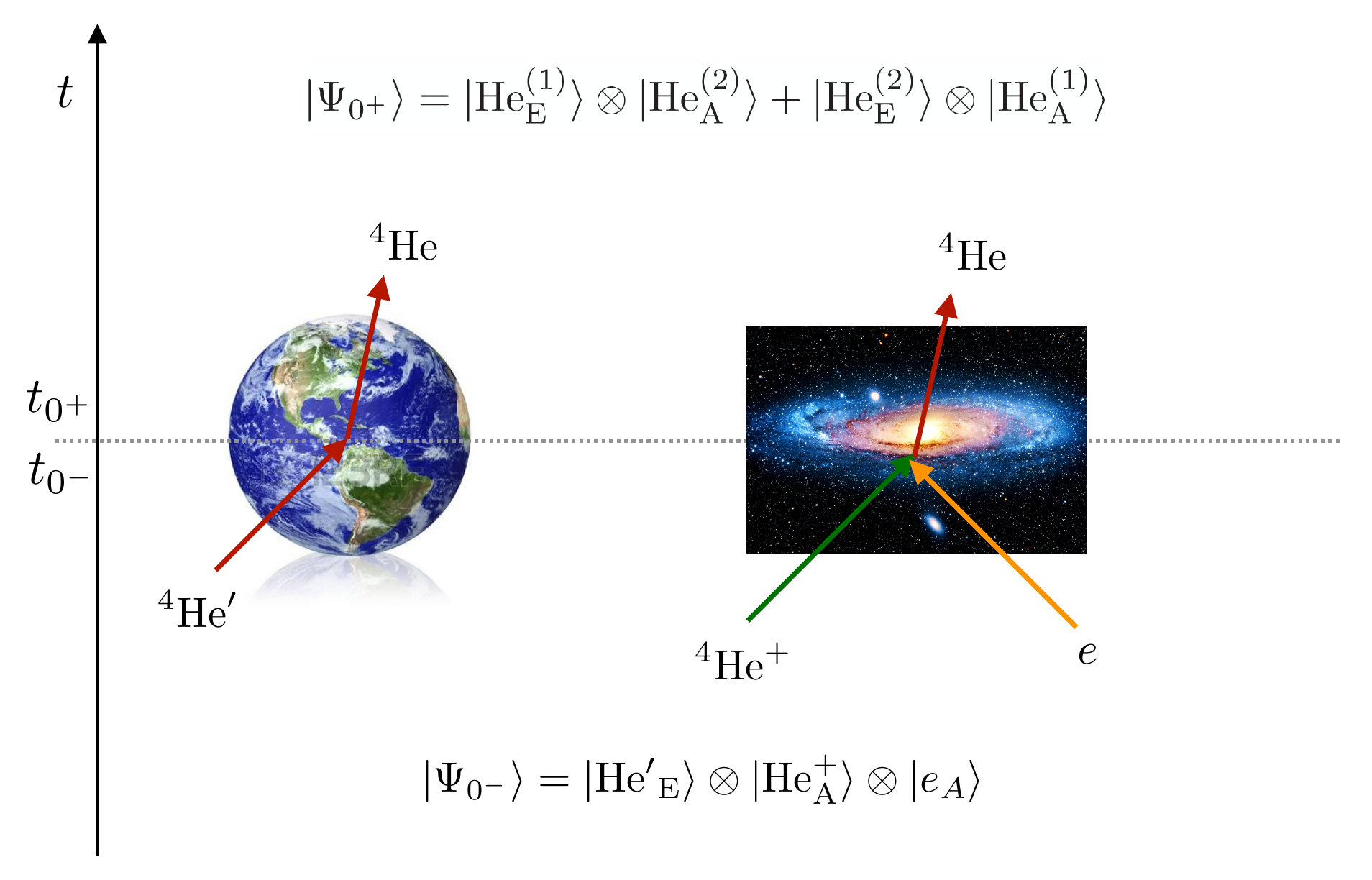}
\caption{Simultaneous generation on the Earth and on the Andromeda galaxy of two identical helium atom states.}
\label{EA}
\end{figure*}

Ordinarily the SA treats identical particles as if they were not and in general it works well.
Here we briefly describe how this method however gives rise to unnecessary methodological and practical difficulties especially when particle quantum correlations are involved. 
These difficulties arise because of the adoption of unphysical labels to mark particles and of the required (anti)symmetrization of states with respect to labels \cite{cohen2005quantum}.

Such a symmetrization implies that, given a set of identical particles, each of them stays with the same probability amplitude in all the occupied single particle states of the system. This, for example, justifies to say that "\textit{only one fermion can occupy a quantum state}" is a not accurate statement \cite{peresbook}.
Other peculiar aspects arise when one considers time dependent problems. In fact, changes performed locally on particles in far away regions instantaneously reflect on all the particles of the system. 

In order to evidence this point, we consider an helium atom $^4\mathrm{He}$ in the state $\ket{\mathrm{He}'_\mathrm{E}}$ on the Earth and the ionized atom $^4\mathrm{He}^{+}$ in the state $\ket{\mathrm{He}^{+}_\mathrm{A}}$ plus an electron in the state $\ket{e_\mathrm{A}}$ on the Andromeda galaxy (subscripts E and A represent the spatial localisation of states on Earth and Andromeda). Being at the beginning all the involved particles distinguishable, the global state is the tensor product $|\Psi_{0^-}\rangle=|\mathrm{He}'_\mathrm{E}\rangle \otimes |\mathrm{He}^+_\mathrm{A}\rangle \otimes |e_\mathrm{A}\rangle$.
At the universal time $t=0$, $^4\mathrm{He}'$ is scattered in $^4\mathrm{He}$ and simultaneously ion $^4\mathrm{He}^+$ on Andromeda absorbs the electron $e$ forming the atom $^4\mathrm{He}$ (see Fig.~\ref{EA}). At $t=0^+$ on the Earth and on Andromeda two identical bosons appear in the states $|\mathrm{He}_\mathrm{E}\rangle$ and $|\mathrm{He}_\mathrm{A}\rangle$. Being in the SA the identical atoms distinguished with labels 1 and 2, the global state is (except a normalisation factor) $|\Psi_{0^+}\rangle=|\mathrm{He}^{(1)}_\mathrm{E}\rangle\otimes |\mathrm{He}^{(2)}_\mathrm{A}\rangle+|\mathrm{He}^{(2)}_\mathrm{E}\rangle\otimes |\mathrm{He}^{(1)}_\mathrm{A}\rangle$. 
Thus, while at $t=0^-$ each particles is separately localised on either E or A, at $t=0^+$ each of the two helium atoms simultaneously occupies both states on E and A. This approach requires to accept the notion, for instance, that the nonrelativistic helium atom generated at $t=0^+$ in A, because of the identical particle in E, instantaneously develops a nonzero amplitude of being there, although the events $(0^+_\mathrm{E},0^+_\mathrm{A})$ are spacelike separated. Moreover the global state of the two identical particles $|\Psi_{0^+}\rangle$ has the form of an entangled state even if they are independently prepared far away from each other. To cope with this situation, that is to avoid observable effects of unobservable labels, "\textit{we must now convince ourselves that this entanglement is not matter of concern}" \cite{peresbook}.

The viewpoint of the SA thus makes it problematic a straightforward discussion of correlations (such as entanglement) in systems of identical particles, because of the difficulty in formally separating the real part of correlations from the unphysical one arising from labels. Moreover, such a description hinders the use of partial trace and the von Neumann entropy, as normally done for nonidentical particles \cite{balachandran2013PRL}. In fact, indistinguishability implies that the particles are not individually addressable and so the common reduced density matrix obtained by partial trace is meaningless. This issue has originated different treatments for a faithful quantification of identical particle entanglement \cite{balachandran2013PRL,plenio2014PRL,ghirardi2004PRA, schliemann2001quantum,eckert2002AnnPhys,wiseman2003PRL, dowlingPRA,vaccaroIJQI,jonesPRA,buscemi2007PRA,vogel2015PRA, giulianoEPJD, benatti2012AnnPhys,sasaki2011PRA,benatti2012PRA, benattiOSID2017,franco2016quantum,sciara2017universality}.

\section{Non-standard approach (NSA) to many identical particles}
\label{Sec:NSA}

%It seems that the idea of treating identical particles as if they were not-identical isn't the best way to proceed and it has been proved it is not even necessary. 

Here we consider a recently introduced approach to deal with identical particles which does not adopt unphysical labels to mark them \cite{franco2016quantum}. This non-standard approach (NSA) eliminates \textit{ab ovo} the conceptual strains inherent in the SA and also allows us to directly focus on the treatment of physical quantum correlations. So far, it has been applied to the case of two identical particles \cite{franco2016quantum,sciara2017universality}. We now re-examine this approach from a fundamental viewpoint and extend it to a system of many identical particles.

Let's take $N$ identical particles, each in a given 1-particle state defined by a complete set of commuting observables. In the following, particle states are characterised by a spatial wavefunction $\phi_k$ and a pseudospin $\sigma_k$, that is $|\phi_k,\sigma_k\rangle:=|k\rangle$. 
The $N$-particle state $|\phi^{(N)}\rangle$, expressed by a complete set of commuting observables is
\begin{equation}\label{psiN}
|\phi^{(N)}\rangle:=|1,2,...,N\rangle
\end{equation}
that does not tell us which particle is in $|k\rangle$ ($k=1,2,...,N$) but simply lists the single-particle states.

In the example illustrated in Fig.~\ref{EA}, being the single-particle states at $t=0^+$ given by $|\mathrm{He}_\mathrm{E}\rangle$ and $|\mathrm{He}_\mathrm{A}\rangle$, using \eqref{psiN} the global final state is $|\mathrm{He}_\mathrm{E},\mathrm{He}_\mathrm{A}\rangle$. This means that one helium atom is localised in E and one in A. Entanglement with respect to labels simply does not appear and the question whether the two identical atoms are entangled is not even needed to be posed.

For one particle, the relevant quantities to get probabilities are the transition amplitudes $\langle k'|k\rangle$ to find the particle in the exit state $|k'\rangle$ if it is initially in the enter state $|k\rangle$. In the NSA, this must be generalised for $N$ particles. The absence of labels implies that the state of equation \eqref{psiN} is not separable in terms of tensor products of 1-particle states and is a holistic entity.

\begin{figure*}[t!]
\centering\includegraphics[width=0.7\textwidth]{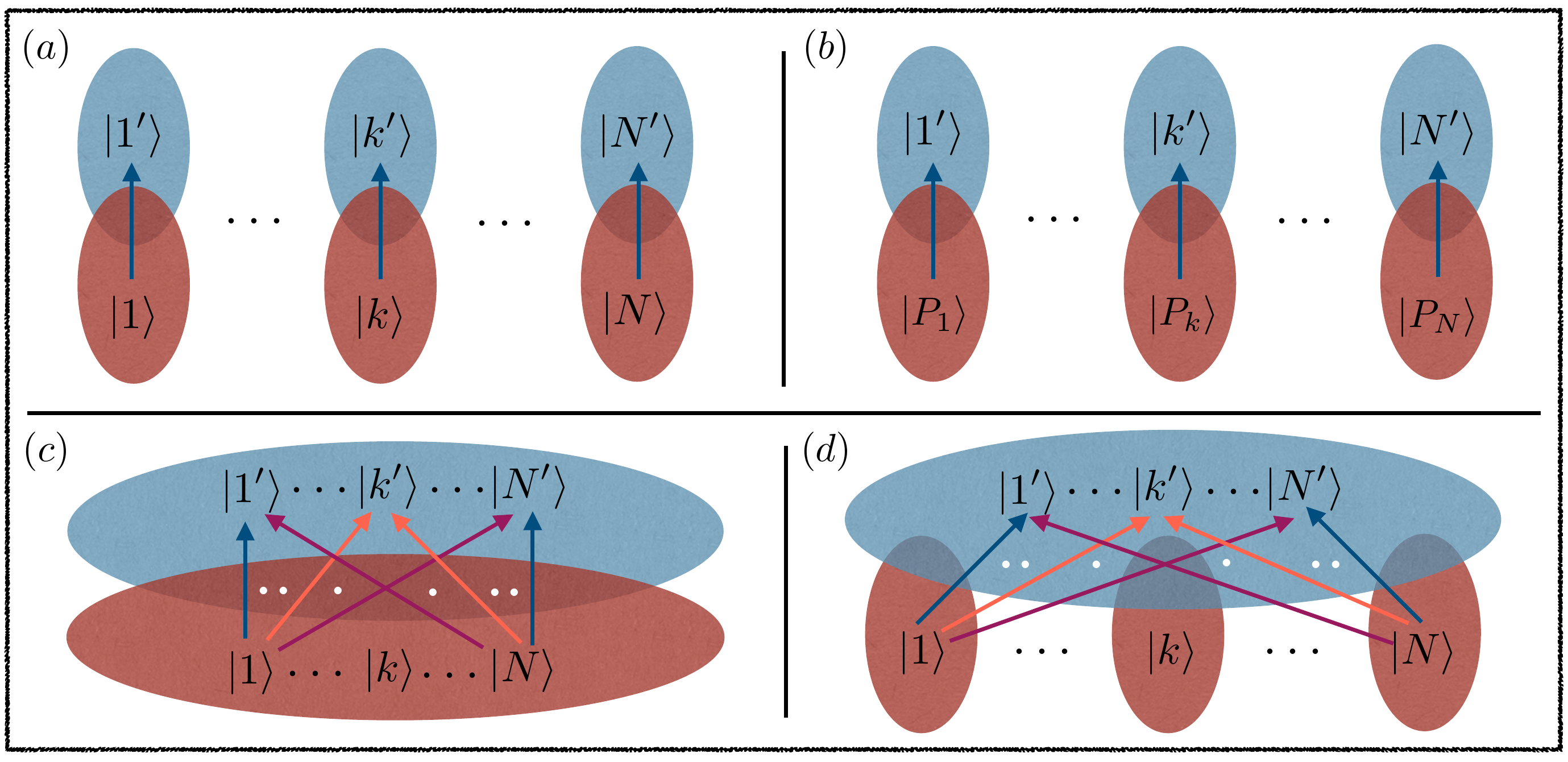}
\caption{(a) and (b) Cluster decomposition principle. The set $\{P_1,...,P_N\}$ represents the $N!$ permutations of the 1-particle states $|1\rangle, ..., |N\rangle$. (c) General case where there is spatial overlap among the enter 1-particle states in which the global system is prepared (red cloud) and among the exit ones on which the system is measured (blue cloud). (d) Particular situation where only all the enter 1-particle states do not overlap (red clouds). The coloured clouds represent the spatial regions occupied by the states written within them. The arrows show the transition of each $|k\rangle$ state ($k=1,...,N$) towards one or more $|k'\rangle$ states ($k'=1',...,N'$).}
\label{clusterN}
\end{figure*}

We assume that the transition amplitude from the state $|1,...,N\rangle$ to the state $|1',...,N'\rangle$, namely $\langle 1',...,N'|1,...,N\rangle$, can be expressed in terms of the 1-particle probability amplitudes. This is natural when each 1-particle state in the transition amplitude is localised in a region far away from the others (see Fig.~\ref{clusterN}(a) and (b)). Under this condition the cluster decomposition principle, stating that distant experiments provide independent outcomes \cite{peresbook}, allows us to express the total transition amplitude as the product of the 1-particle ones as $\langle 1',2',...,N'|1,2,...,N\rangle :=\langle 1'|1\rangle\langle 2'|2\rangle...\langle N'|N\rangle$ (see Fig.\ref{clusterN}(a)), while more generally as

\begin{equation}\label{permut}
\langle 1',2',...,N'|1,2,...,N\rangle :=\langle 1'|P_1\rangle\langle 2'|P_2\rangle...\langle N'|P_N\rangle,
\end{equation}
where the set $P_1,P_2,...,P_N$ represents the case in which the state $|P_k\rangle$ occupies the $k$-th region (see Fig.~\ref{clusterN}(b)). The transition amplitude \eqref{permut} is linear in each of the 1-particle states.

When both the enter and exit 1-particle states are localised in overlapping spatial regions (see Fig.~\ref{clusterN}(c)), to maintain the property of linearity, the $N$-particle probability amplitude can be expressed as a linear combination of $N!$ terms of the form \eqref{permut}
\begin{align}\label{amplN}
\begin{split}
&\langle 1',2',...,N'|1,2,...,N\rangle:=\\
&\sum_P \alpha_P\langle 1'|P_1\rangle \langle 2'|P_2\rangle ...\langle N'|P_N\rangle,
\end{split}
\end{align}
where $P=\{P_1,P_2,...,P_N\}$ runs over all the 1-particle state permutations. Taking into account that 
\begin{equation}\label{cond}
\langle 1',2',...,N'|1,2,...,N\rangle^{\ast}=\langle 1,2,...,N|1',2',...,N'\rangle,
\end{equation}
we find $\alpha_P^{\ast}=\alpha_P$, so the expansion \eqref{amplN} and the condition \eqref{cond} are consistent only if the coefficients $\alpha_P$ are real. We now consider the simplest case of 2 identical particles, for which the 2-particle probability amplitude, using \eqref{amplN}, is
\begin{equation}\label{2amplcoef}
\langle 1',2'|1,2\rangle:=a\langle 1'|1\rangle \langle 2' |2\rangle+b \langle 1' |2\rangle\langle 2' |1\rangle.
\end{equation}
The equation \eqref{2amplcoef} associates the order of the states in the 2-particle probability amplitude on the left with the order of the products of the 1-particle probability amplitudes on the right. Swapping the single-particle states in the 2-particle state vector exchanges the weights of the single-particle products. However, this swapping cannot modify the 2-particle amplitude implying that amplitudes may differ only for a global phase factor, that is
\begin{align}
\begin{split}
\langle 1',2'|2,1\rangle &=e^{i\zeta}\langle 1',2'|1,2\rangle\\
&=e^{i\zeta}a\langle 1'|1\rangle \langle 2'|2\rangle+e^{i\zeta}b\langle1'|2\rangle \langle 2'|1\rangle\\
&=a\langle 1'|2\rangle \langle 2'|1\rangle+b \langle 1'|1\rangle \langle 2'|2\rangle.
\end{split}
\end{align}
We find that $e^{i\zeta}b=a $ and $e^{i\zeta}a=b$, so $|a|=|b|$ and $(e^{i\zeta})^2:=\eta^2=1$ from which $\eta=\pm 1$. In this way, taking $a=1$ in the linear combination of equation \eqref{2amplcoef}, $b=\eta$. So, the inner product of two "holistic" state vectors is
\begin{equation}\label{2systampleta}
\langle 1',2'|1,2\rangle:=\langle 1' |1\rangle \langle 2' |2\rangle+\eta \langle 1' |2\rangle\langle 2' |1\rangle
\end{equation}
which represents the core of our approach and includes the particle spin-statistics principle. In fact, according to the Pauli exclusion principle, the probability amplitude of finding two fermions in the same state is zero, so $\langle 1',1'|1,2\rangle=(1+\eta)\langle 1'|1\rangle \langle 1'|2\rangle$ requires that $\eta$ is $-1$. The choice $\eta=+1$ gives the maximum amplitude of finding two particles in the same state and corresponds to the case of bosons. 

Generalising the symmetric and antisymmetric expression \eqref{2systampleta} to an arbitrary number $N$ of identical particles, we can write
\begin{align*}\label{Nampleta}
\begin{split}
&\langle 1',2',...,N'|1,2,...,N\rangle :=\\
&\sum_P\eta^P\langle 1'|P_1\rangle\langle 2'|P_2\rangle...\langle N'|P_N\rangle,
\end{split}
\end{align*}
where, in analogy with the 2-particle case, for bosons $\eta^P$ ($P$ being the parity of the permutation) is always 1 and for fermions it is $1$ ($-1$) for even (odd) permutations. Equation \eqref{Nampleta} induces the symmetrization property of the $N$-system state space: $|1,2,...,j,...,k,...,N\rangle=\eta|1,2,...,k,...,j,...,N\rangle$, for $j,k=1,...,N$. Linearity of the $N$-system state vector with respect to each 1-particle state immediately follows from the linearity of the 1-particle amplitudes and the $N$-particle state vectors thus span the physical symmetric state space $\mathcal{H}_{\eta}^{(N)}$.

We remark that in the situations represented by Fig.~\ref{clusterN}(b) and only in this case one can write $\langle 1',...,N'|P_1,...,P_N\rangle :=(\langle 1'| \otimes ...\otimes \langle N'|)(|P_1\rangle \otimes...\otimes |P_N\rangle)$. Therefore, for calculation purposes $|1,...,N\rangle \simeq|1\rangle \otimes ... \otimes |N\rangle$. This is not true if there is overlap either among the 1-particle enter states or among the exit ones (see Fig.\ref{clusterN}(c) and (d)). In this sense probability amplitudes are more fundamental than quantum states.

An arbitrary elementary normalised $N$-identical particle state is defined as
\begin{equation}\label{PsiNnorm}
|\Phi^{(N)}\rangle:=\dfrac{1}{\mathcal{N}}|\phi^{(N)}\rangle :=\dfrac{1}{\mathcal{N}}|1,2,...,N\rangle,
\end{equation}
where $\mathcal{N}=\sqrt{\langle 1,2,...,N|1,2,...,N\rangle}$.
$|\Phi^{(N)}\rangle$ is expressed in terms of single-particle states as a single state vector, which is to be compared with the $N!$ product state vector in the SA approach: $|\Theta^{(N)}\rangle=\dfrac{1}{\mathcal{N}}\sum_P \eta^P|1_{P_1}\rangle \otimes |2_{P_2}\rangle \otimes ... \otimes |N_{P_N}\rangle$.

The next step is to define the action of operators in the NSA approach. We limit ourselves to an arbitrary 1-particle operator $A^{(1)}$ that acts on each 1-particle state at a time: $A^{(1)}|k\rangle:=|A^{(1)}k\rangle$. Its action on $N$-particle states is naturally defined as 
\begin{equation}\label{A1}
A^{(1)}|1,2,...,N\rangle := \sum_k|1,...,A^{(1)}k,...,N\rangle.
\end{equation}

\subsection{Partial trace}
To calculate the partial trace we have to define a product operation between bra and ket of different number of particles. In particular, the $M$-particle partial trace of an $N$-particle state is expressed as

\begin{equation}\label{trM}
\mathrm{Tr}^{(M)}|\Phi^{(N)}\rangle\langle \Phi^{(N)}|:=\sum_{\tilde{k}'}\langle \tilde{k}'|\cdot |\Phi^{(N)}\rangle\langle \Phi^{(N)}|\cdot |\tilde{k}'\rangle,
\end{equation}
where $\{|\tilde{k}'\rangle:=\ket{k'_1,\ldots,k'_M}/\mathcal{N}_{k'}\}$ is a collective $M$-particle orthonormal basis ($\mathcal{N}_{k'}$ being a normalisation constant). This operation corresponds to measure the states of $M$ identical particles without registering the outcomes and for $M=N$, it just coincides with the normalised probability amplitude. 

We now consider the simple case in which $M=1$.
Given a single-particle orthonormal basis $\{\ket{k'}\}$, we have to define the product $\langle k'|\cdot|1,2,...,N\rangle$. Given the operator $A^{(1)}=|j'\rangle \langle k'|$ and using equation \eqref{A1}, its action on a $N$-particle state $|1,...,N\rangle$ can be written as 
\begin{align}\label{operatoractionN}
\begin{split}
A^{(1)}|1,...,N\rangle&:=\sum_k|  
1,2,...,j' \langle k'|k\rangle,...,N\rangle \\
&=\sum_{k}\eta^{k-1}\langle k'| k\rangle |  
j',1,2,...,\cancel{k},...,N\rangle,
\end{split}
\end{align} 
where, in the second line, we have taken out of the $N$-particle ket the complex number $\langle k'|k\rangle$ and shifted the state $|j'\rangle$ from the $k$-th site to the first one ($\cancel{k}$ is an elegant notation to indicate the lack of $k$-th state!). Let's express now the ket in the second line of equation \eqref{operatoractionN} as follows: $ |  
j',1,2,...,\cancel{k},...,N\rangle := |j'\rangle \wedge  |1,2,...,\cancel{k},...,N\rangle$, where we have introduced with the symbol $\wedge$ a non-separable symmetric external product between different kets, valid for boson and fermions, that we call \textit{wedge} product (for fermions this product is the Penrose's wedge product defined in terms of labelled states \cite{penrosebook}). For $N$ identical particles we have $|1,2,...,N\rangle:=|1\rangle \wedge |2\rangle \wedge...\wedge |N\rangle$ and $(|1,2,...,N\rangle)^{\dagger}:=\langle N|\wedge...\wedge \langle 2| \wedge \langle 1|$. Moreover, $|P_1\rangle \wedge |P_2\rangle \wedge... \wedge |P_N\rangle=\eta^P|1\rangle \wedge |2\rangle \wedge ... \wedge |N\rangle$. The wedge product coincides with the multiplication operation of the exterior algebra associated to the symmetrized Hilbert space $\mathcal{H}_{\eta}^{(N)}$.  Therefore, equation \eqref{operatoractionN} can be written as
\begin{equation}\label{wedge}
A^{(1)}|1,...,N\rangle:=|j'\rangle \wedge \sum_k\eta^{k-1}\langle k'| k\rangle |1,2,...,\cancel{k},...,N\rangle,
\end{equation}
and it suggests the introduction of a generalised \textit{dot} product operation between bra and ket of different dimensionality $M$ and $N$ respectively, which in the case of $M=1$ is
\begin{equation}\label{ldot}
\langle k'|\cdot|  1,2,...,N\rangle :=\sum_{k=1}^N\eta^{k-1}\langle k'|k\rangle|1,...,\cancel{k},...,N\rangle.
\end{equation} 
The equation \eqref{ldot} defines the projection of a $N$-particle state on a 1-particle state and gives an (unnormalised) $(N-1)$-particle state. 
We can see that if the $N$-particle state is explicitly expressed in terms of wedge products, the dot product is distributive with respect to the wedge.

By taking the 1-particle projection operator $\Pi^{(1)}_{k'}=|k'\rangle\langle k'|$, the probability $p_{k'}$ of finding one of the identical particles in the state $|k'\rangle$ is given (except a normalisation factor) by $\langle \Pi_{k'}^{(1)}\rangle_{\Phi^{(N)}}$. Being the action of the 1-particle identity operator $\mathbb{I}^{(1)}=\sum_{k'}\Pi_{k'}^{(1)}$ on a $N$-particle state $\mathbb{I}^{(1)}|1,...,N\rangle=N|1,...,N\rangle$, the normalised reduced $(N-1)$-particle pure state $|\Phi_{k'}^{(N-1)}\rangle$ and the probability $p_{k'}$ are
\begin{equation}
 |\Phi_{k'}^{(N-1)}\rangle=\dfrac{\langle k' |\cdot |\Phi^{(N)}\rangle}{\sqrt{\langle \Pi_{k'}^{(1)}\rangle_{\Phi^{(N)}}}},\quad
 p_{k'}=\dfrac{\langle \Pi_{k'}^{(1)}\rangle_{\Phi^{(N)}}}{\langle\mathbb{I}^{(1)}\rangle_{\Phi^{(N)}}}.
\end{equation}
If all the 1-particle states in $|\Phi^{(N)}\rangle$ are orthonormal, i.e. $\langle i|j\rangle=\delta_{ij}$, one has $p_{k'}=\sum_{j=1}^N\dfrac{|\langle k'|j\rangle|^2}{\langle \mathbb{I}^{(1)}\rangle_{\Phi^{(N)}}}$ which corresponds to the sum of probabilities of incompatible outcomes.

Specifically the normalised 1-particle partial trace of an $N$-particle state is

\begin{align}\label{partialtraceN}
\begin{split}
\rho^{(N-1)} & := \frac{1}{\langle\mathbb{I}^{(1)}\rangle_{\Phi^{(N)}}}\mathrm{Tr}^{(1)}|\Phi^{(N)}\rangle\langle\Phi^{(N)}|\\
& = \sum_{k'} p_{k'} |\Phi^{(N-1)}_{k'}\rangle \langle\Phi^{(N-1)}_{k'}|.
\end{split}  
\end{align}

We now extend the above treatments to the case $M=2$ and $|\tilde{k}'\rangle :=|l',m'\rangle/\mathcal{N}_k$ (because it is useful in the following). The relevant dot product is
\begin{align}\label{dot2}
\begin{split}
&\langle l',m'|\cdot |1,...,N\rangle=\\
&\sum_{\substack{j,k=1 \\ k\neq j}}^N \eta^{j+k-(1+2)}\langle l'|j\rangle\langle m'|k\rangle |1,...,\cancel{j},\cancel{k},...N\rangle,
\end{split} 
\end{align}
and the $2$-particle unity operator is $\mathbb{I}^{(2)}=(1/2!)\sum_{\tilde{k}'} \Pi_{\tilde{k}'}^{(2)}$, where $\Pi_{k'}^{(2)}=|\tilde{k}'\rangle\langle\tilde{k}'|$ is the $2$-particle projection operator.
The normalised reduced $(N-2)$-particle pure state $|\Phi^{(N-2)}_{\tilde{k}'}\rangle$ and its probability $p_{\tilde{k}'}$ are
\begin{equation}\label{projectionM}
|\Phi^{(N-2)}_{\tilde{k}'}\rangle=
\frac{\langle \tilde{k}'|\cdot |\Phi^{(N)}\rangle}{\sqrt{\langle\Pi_{\tilde{k}'}^{(2)}\rangle_{\Phi^{(N)}}}}, \quad
p_{\tilde{k}'}=\dfrac{1}{2!}\dfrac{\langle\Pi_{\tilde{k}'}^{(2)}\rangle_{\Phi^{(N)}}}{\langle\mathbb{I}^{(2)}\rangle_{\Phi^{(N)}}}.
\end{equation}
The $(N-2)$-particle reduced density matrix is 
\begin{align}
\label{partialtraceM}
\begin{split}
\rho^{(N-2)}&:=\frac{1}{2!\langle\mathbb{I}^{(2)}\rangle_{\Phi^{(N)}}}\mathrm{Tr}^{(2)}|\Phi^{(N)}\rangle\langle{\Phi^{(N)}}|\\
&=\sum_{\tilde{k}'} p_{\tilde{k}'} |\Phi^{(N-2)}_{\tilde{k}'}\rangle\langle\Phi^{(N-2)}_{\tilde{k}'}|.
\end{split}  
\end{align}
When required, the generalisation of equation \eqref{dot2} to the case of any $M$ is straightforward.

We point out that, in the NSA, once the relevant reduced density matrix is obtained, the entanglement between the bipartition of $M$ and $(N-M)$ particles can be measured by the von Neumann entropy $S(\rho^{(N-M)})=-\mathrm{Tr}(\rho^{(N-M)}\log_2\rho^{(N-M)})$. Moreover, knowledge of the reduced density matrices of all the possible bipartitions of the system allows the qualitative assessment of the genuine multipartite entanglement of the pure state of $N$ identical particles, as usually done for nonidentical particles \cite{horodecki2009quantum}. This result differs from what is obtained in the SA where the partial trace cannot be used \cite{balachandran2013PRL}. We shall analyse identical particle entanglement in Sec~\ref{Sec:App}.

\subsection{Connection of NSA with second quantization}
Above we have introduced relations between states differing in the number of particles. This suggests a relationship between the NSA to identical particles and second quantization. However, while in the second quantization particles are elementary excitations of fields, our approach applies to any system of identical quantum objects. In the second quantization approach creation and annihilation operators connect states differing only by one particle and their commutation rules reflect the commutation rules of the fields. In the NSA the dot product connect states differing for a generical number of particles. Finally, in the second quantization the foundamental role is played by the commutation rules, while in the NSA it is played by the symmetry of the states.

To show the connection between the NSA and the second quantization, we notice that the equation \eqref{ldot} suggests the introduction of a 1-particle annihilation operator
\begin{equation}\label{a}
a(k)|1,...,N\rangle:=\langle k|\cdot|1,...,N\rangle
\end{equation}
and its the adjoint is
\begin{equation}\label{adag}
a^{\dagger}(k)|1,...,N\rangle:=|k\rangle \wedge |1,...,N\rangle.
\end{equation}
The equation \eqref{dot2} for example directly suggests the introduction of two-states annihilation and creation operators defined as 
\begin{equation}\label{aa}
a(j,k)|1,...,N\rangle:=\langle j,k| \cdot | 1,...,N\rangle,
\end{equation}
\begin{equation}\label{aadag}
a^{\dagger}(j,k)|1,...,N\rangle:=|j,k\rangle \wedge | 1,...,N\rangle.
\end{equation}
Operators $a(j,k)$ and $a^{\dagger}(j,k)$  annihilate and create couples of particles and, as a consequence of the holistic nature of the states in the NSA, $a(j,k)\neq a(j)a(k)$.

From the symmetry properties of the states, it is simple to obtain the $a$'s commutation rules. For the 1-particle operators in equations \eqref{a} and \eqref{adag} we obtain
\begin{align}
\begin{split}
&[a(j),a^{\dagger}(k)]_{\eta}=\langle j|k\rangle \\
&[a^{\dagger}(j),a^{\dagger}(k)]_{\eta}=[a(j),a(k)]_{\eta}=0,
\end{split}
\end{align}
where $[a(j),a^{\dagger}(k)]_{\eta}=a(j)a^{\dagger}(k)-\eta a^{\dagger}(k)a(j)$ is a commutator for bosons and an anticommutator for fermions. 
Instead, from the 2-particle operators in equations \eqref{aa} and \eqref{aadag} one always obtains commutation rules even if they create or annihilate fermionic states
\begin{align}\label{comm2}
\begin{split}
&[a(j,k),a^{\dagger}(m,n)]=\langle j,k|m,n\rangle \\
&[a^{\dagger}(j,k),a^{\dagger}(m,n)]=[a(j,k),a(m,n)]=0.
\end{split}
\end{align}
In the first equation of \eqref{comm2} the right side keeps the memory of the bosonic or fermionic nature of the 1-particle states with the presence of $\eta$ in the expansion of $\langle j,k|m,n\rangle$ in terms of the 1-particle probability amplitudes.

\section{Properties of some multi-particle states}
\label{Sec:GenStates}

In this Section we show how some known structures and properties of states of nonidentical particles can be generalised to the case of identical particles.

\subsection{Spin exchanged states}

We consider the well-known W entangled state of nonidentical particles, which constitutes an important resource state for several quantum information tasks \cite{eisertReview,BLFC2017}. We explicitly describe a single-particle state with the spatial mode $\phi_i$ ($i=1,\dots,N$) and the pseudospin $\uparrow, \downarrow$. The W state has the structure
\begin{align} \label{noW}
\begin{split}
|W\rangle=&\ |\phi_1 \uparrow,\phi_2 \downarrow,...,\phi_N \downarrow\rangle+|\phi_1 \downarrow,\phi_2 \uparrow,...,\phi_N \downarrow\rangle\\
&+...+|\phi_1 \downarrow,...,\phi_{N-1} \downarrow,\phi_N \uparrow\rangle, 
\end{split}
\end{align}
where for nonidentical particles each term of the superposition is a tensor product of single-particle states. The very same structure is considered to hold also for identical particles in separated locations. In this case, is this structure valid under any circumstance? To analyse this aspect, we choose 3 fermions placed in three separated spatial modes $A$, $B$ and $C$, so that the W state given in equation \eqref{noW} is $|W_3\rangle=|A\uparrow,B \downarrow,C \downarrow\rangle+|A\downarrow, B \uparrow,C \downarrow\rangle+|A\downarrow, B\downarrow,C \uparrow\rangle$. When two fermions are in the same spatial mode, for instance $A=B$, this state becomes
\begin{equation}
|\bar{W}_3\rangle=|A \uparrow,A \downarrow, C \downarrow\rangle+|A \downarrow,A \uparrow, C \downarrow\rangle,
\end{equation}
where the state $|A \downarrow,A \downarrow, C \uparrow\rangle$ does not appear because of the Pauli exclusion principle. Using the symmetry properties of the elementary states, $|A \downarrow,A \uparrow, C \downarrow\rangle=- |A \uparrow,A \downarrow, C \downarrow\rangle$, one gets $|\bar{W}_3\rangle=0$, while one expects no restriction arising from particle statistics, considering that the two fermions in $A$ have opposite spins. The same problem is thus expected to arise even when the particles are spatially nonoverlapping but non-local measurements are performed. Hence, the form of the W state of equation \eqref{noW} for identical particles does not work in general.

We define a spin exchanged state (SPES) as a suitable linear combination of elementary states where only particle pseudospins are exchanged, that is
\begin{equation}\label{SPES}
|\mathrm{SPES}\rangle=\dfrac{1}{\mathcal{N}}\sum_{P}\dfrac{\eta^{P}}{\mathcal{N}_P}|\phi_1 \ \sigma_{P_1},...,\phi_N \ \sigma_{P_N}\rangle,
\end{equation}
where $P$ are here the cyclic permutations of pseudospins $\sigma_{P_i}$, $\mathcal{N}_P$ is the normalisation constant of each state entering the sum and $\mathcal{N}$ is the global normalisation constant. When $\sigma_{1}=\ \uparrow$ and $\sigma_2=\ldots=\sigma_N=\ \downarrow$, the SPES represents a generalisation of the W state for identical particles valid bosons and fermions under any circumstances. Other states of interest are obtained from \eqref{SPES} for different pseudospin conditions; for instance, assuming a ring configuration of the $N$ particle and taking $\sigma_{1}=\ldots=\sigma_M=\ \uparrow$ and $\sigma_{(M+1)}=\ldots=\sigma_N=\ \downarrow$ such states can represent a linear combination of spin block systems in the ring chain.

\subsection{Separability of spatial and spin degrees of freedom}
For any elementary state of nonidentical particles, the spatial and spin degrees of freedom, being associated to the individual particle, are independent and separable. This is not the case for elementary states of identical particles. 

As mentioned above, the essence of the NSA is the absence of labels and the consequent holistic form of the states that are not separable in tensor products of single-particle states. 
For a single particle, we have $|\phi \ \sigma\rangle \equiv |\phi\rangle \otimes |\sigma\rangle$, i.e. the spatial part of the state can be treated separately from the pseudospin one. Instead, for two identical particles, when $\phi_1\neq \phi_2$ and $\sigma_1\neq\sigma_2$, it is $|\phi_1 \, \sigma_1,\phi_2 \, \sigma_2\rangle \neq |\phi_1,\phi_2\rangle \otimes |\sigma_1, \sigma_2\rangle$, so measures of position and pseudospin operators may be not independent. One may ask which structure a state of identical particles must have in order that spatial and pseudospin degrees of freedom are independent from each other. 

To fix our ideas, let us take the state
\begin{equation}\label{sep}
|\phi\rangle=\alpha|\varphi_1 \, \sigma,\varphi_2 \, \tau\rangle+\beta |\varphi_1 \, \tau,\varphi_2 \, \sigma\rangle,
\end{equation}
with $\varphi_1 \neq \varphi_2$, $\tau\neq \sigma$ and calculate the probability amplitude on the state $|\phi'\rangle=|\varphi'_1 \ \sigma',\varphi'_2 \ \tau'\rangle$, that is
\begin{equation}\label{amplitude}
\langle \phi'|\phi\rangle:=\langle\varphi'_1 \, \sigma',\varphi'_2 \, \tau'|\{\alpha|\varphi_1 \, \sigma,\varphi_2 \, \tau\rangle+\beta |\varphi_1 \, \tau,\varphi_2 \, \sigma\rangle \}.
\end{equation}
Using the expression \eqref{2systampleta}, we find that only when $\alpha=1$ and $\beta=\pm 1$ equation \eqref{amplitude} can be written as
\begin{equation}
\langle \phi'|\phi\rangle=\langle \varphi'_1  \varphi'_2|\varphi_1  \varphi_2\rangle_{\eta\beta}\langle\sigma'\tau'|\sigma \tau\rangle_{\beta},
\end{equation}
where $\langle \mu' \ \nu' | \mu \ \nu \rangle_{\gamma}$ indicates the probability amplitude of the form of equation~\eqref{2systampleta} with $\eta$ substituted by $\gamma$.
We have thus shown that, under the above conditions, in the entangled states of the form $|\Phi^{\pm}\rangle=\dfrac{1}{\mathcal{N}}(|\varphi_1 \, \sigma,\varphi_2 \, \tau\rangle\pm |\varphi_1 \, \tau, \varphi_2 \, \sigma\rangle)$, spatial and pseudospin degrees of freedom can be separated as $|\Phi^{\pm}\rangle=|\varphi_1, \varphi_2 \rangle_{\eta\beta} \otimes |\sigma,\tau\rangle_{\beta}$, where the subscripts indicate the symmetry of the state: $|\kappa,\chi\rangle_{\gamma}=\gamma|\chi,\kappa\rangle_\gamma$. For such a state, entanglement of pseudospins can be treated independently of the spatial degrees of freedom. When $\sigma=\uparrow$ and $\tau=\downarrow$, $|\Phi^{\pm}\rangle$ are two Bell states.  
It is moreover immediate to show that an entangled state of the form 
$|\Psi\rangle=\dfrac{1}{\mathcal{N}}(|\phi_1 \ \sigma,\phi_2 \ \sigma\rangle\pm |\phi_1 \ \tau, \phi_2 \ \tau\rangle)$ is always equivalent to $|\phi_1\phi_2\rangle_{\eta}\otimes(|\sigma,\sigma\rangle+|\tau,\tau\rangle)$.

\section{Applications}
\label{Sec:App}

We now apply the NSA formalism and tools described above to examine entanglement properties of identical particle pure states by partial traces and local measurements.

\subsection{Effects of the spatial overlap on the entanglement in SPES}

We consider the SPES state for 3 qubits placed in separated spatial modes $L$ (left), $C$ (center) and $R$ (right). Following equation \eqref{SPES}, it has the form
\begin{align}\label{W3}
\begin{split}
|\mathrm{SPES}_3\rangle=&\dfrac{1}{\sqrt{3}}(|L\uparrow,C \downarrow, R \downarrow\rangle+\eta|L\downarrow,C \uparrow, R \downarrow\rangle\\
&+|L\downarrow,C \downarrow, R \uparrow\rangle).
\end{split}
\end{align}
Partially tracing this state over the one-particle basis localised in $L$,
$\{\ket{L \uparrow}, \ket{L \downarrow}\}$, we obtain the reduced density matrix
\begin{align}
\begin{split}
\rho_L^{(2)}&=\dfrac{1}{3}(|C\downarrow,R\uparrow\rangle\langle C\downarrow,R\uparrow|+|C\uparrow,R\downarrow\rangle\langle C\uparrow,R\downarrow|\\
&+|C\downarrow,R\downarrow\rangle \langle C\downarrow,R\downarrow|)\\
&+\dfrac{\eta}{3}(|C\downarrow,R\uparrow\rangle \langle C \uparrow,R\downarrow|+|C\uparrow,R\downarrow\rangle\langle C\downarrow,R\uparrow|).
\end{split}
\end{align}
The von Neumann entropy $E_L(\mathrm{SPES}_3)=-\mathrm{Tr}(\rho_L^{(2)}\log_2\rho_L^{(2)})=\log_23-2/3$ measures the entanglement of pseudospins between the bipartitions $L$ (one particle) and $C$-$R$ (two particles). This entropy is less than 1 and independent of the type of particle. As expected, this result coincides with that obtained, under the same conditions, for nonidentical particle W state.

We now consider the case when the spatial wave functions of two particles completely overlap, in particular $C=L$. From equation \eqref{SPES}, the corresponding (unnormalised) SPES is
\begin{equation}
|\mathrm{SPES}'_3\rangle= 2|L\uparrow,L\downarrow,R \downarrow\rangle
+\frac{(1+\eta)}{2} \dfrac{|L\downarrow,L\downarrow,R\uparrow\rangle}{\sqrt{2}}.
\end{equation}
Tracing again over the 1-particle basis in $L$, we find
\begin{align}
\begin{split}
\rho^{(2)}_L&=\dfrac{1}{\mathcal{N}}[4(|L\downarrow,R\downarrow\rangle \langle L\downarrow,R\downarrow|+|L\uparrow,R\downarrow\rangle\langle L\uparrow, R\downarrow|)\\
&+2\eta\kappa(|L\uparrow,R\downarrow\rangle\langle L\downarrow,R\uparrow|+|L\downarrow,R\uparrow\rangle\langle L\uparrow,R\downarrow|)\\
&+\kappa^2|L\downarrow,R\uparrow\rangle\langle L\downarrow,R\uparrow|],
\end{split}
\end{align}
where $\kappa=(1+\eta)^2/(2\sqrt{2})$ and $\mathcal{N}=(8+\kappa^2)$. The corresponding von Neumann entropy is
\begin{equation}\label{EWL}
E(\mathrm{SPES}'_3)=-\dfrac{4}{8+\kappa^2}\log_2\dfrac{4}{8+\kappa^2}-\dfrac{4+\kappa^2}{8+\kappa^2}\log_2\dfrac{4+\kappa^2}{8+\kappa^2}.
\end{equation}
This entropy is for fermions $E_f(\mathrm{SPES}'_3)=1$ and for bosons $E_b(\mathrm{SPES}'_3)=\log_25-\dfrac{3}{5}\log_23-\dfrac{2}{5}$. This result highlights the effect of spatial overlap and of the nature of the particles on the bipartite entanglement of three identical particles.

\subsection{Bell inequality violation for identical particles}
We now apply the NSA to study a state of identical particles within a Bell test scenario \cite{brunner2014publisher,werner1989quantum}, using a suitable operational framework \cite{LFRoperationalframe}. 

We take two independently prepared identical qubits, one being in the spatial mode $\psi$ with pseudospin $\uparrow$ and the other one in the spatial mode $\psi'$ with pseudospin $\downarrow$. The global state is therefore 
\begin{equation}\label{stateOF}
|\Psi\rangle=|\psi \uparrow,\psi' \downarrow \rangle .
\end{equation}
We notice that the configuration described by this state for nonidentical particles does not present entanglement. To make it emerge the entanglement within this state, it is natural to choose local measurements of single-particle pseudospin states performed in two separated restricted spatial regions. This \textit{modus operandi} defines an operational framework founded on spatially localised operations and classical communication (sLOCC), where "spatially localised" pinpoints that we do not refer to a given particle, which is individually unaddressable, but to a given spatial location \cite{LFRoperationalframe}. 

In this context, it is useful to introduce the probability amplitudes of finding the two particles in the two separated spatial regions $L$ (left) and $R$ (right), $l=\langle L|\psi\rangle$, $l'=\langle L|\psi'\rangle$, $r=\langle R|\psi\rangle$ and $r'=\langle R|\psi'\rangle$. Following Ref.~\cite{LFRoperationalframe}, we choose the two-particle basis in the subspace defined by the two separated regions, namely $B_{LR}=\{|L \uparrow,R \uparrow\rangle, |L\uparrow,R \downarrow\rangle, |L\downarrow,R \uparrow\rangle, |L\downarrow,R \downarrow\rangle \}$, we project the state \eqref{stateOF} onto this subspace by means of the projector $\Pi_{LR}=\sum_{\sigma,\tau=\uparrow,\downarrow}|L \sigma,R \tau\rangle \langle L \sigma, R \tau |$ obtaining the pure state
\begin{equation}\label{rhoLR}
\rho_{LR}=|\Psi_{LR}\rangle\langle \Psi_{LR}|,
\end{equation}
where $|\Psi_{LR}\rangle:=\Pi_{LR}|\Psi\rangle/\sqrt{P_{LR}}$ is
\begin{equation}\label{PsiRL}
|\Psi_{LR}\rangle=\dfrac{lr'|L\uparrow,R\downarrow\rangle+\eta l'r|L\downarrow,R\uparrow\rangle}{\sqrt{P_{LR}}},
\end{equation}
with $P_{LR}=\langle \Psi|\Pi_{LR}|\Psi\rangle=|lr'|^2+|l'r|^2$ being the probability of obtaining it. 

We now use the pseudospin observable $\mathcal{O}_S:=\mathbf{O}_S\cdot \bm{\sigma}_S$ ($S=L,R$) with eigenvalues $\pm 1$, where $\mathbf{O}_S$ is the unit vector in an arbitrary direction in the spin space and $\bm{\sigma}_S=(\sigma_x^S,\sigma_y^S,\sigma_z^S)$ is the Pauli matrices vector. The CHSH-Bell inequality in this context is 
\begin{equation}\label{CHSHbell}
\mathcal{B}(\rho_{LR})=|\langle\mathcal{O}_L\mathcal{O}_R\rangle+\langle\mathcal{O}_L\mathcal{O}'_R\rangle+\langle\mathcal{O}'_L\mathcal{O}_R\rangle-\langle\mathcal{O}'_L\mathcal{O}'_R\rangle |\leq2,
\end{equation}
where $\mathcal{B}(\rho_{LR})$ is the Bell function expressed in terms of the correlation functions of the pseudospin observables and $\mathcal{O}'_S$ indicates the measurement in a direction different from that of $\mathcal{O}_S$  \cite{LoFranco2016}. A well-known procedure \cite{horodecki1995violating} allows us to express the maximum value of the Bell function $\mathcal{B}_\mathrm{max}$ in terms of the concurrence of the state $|\Psi_{LR}\rangle$ as
\begin{equation}\label{bmaxC}
\mathcal{B}_\mathrm{max}=2\sqrt{1+C(\Psi_{LR})^2}=2\sqrt{1+\left(\dfrac{2|lr'l'r|}{P_{LR}}\right)^2},
\end{equation}
where the explicit expression of $C(\Psi_{LR})$ has been used.
We have obtained $\mathcal{B}_\mathrm{max}(\rho_{LR})>2$ when $C(\Psi_{LR})>0$, i.e. whenever there is spatial overlap between $\psi$ and $\psi'$ and the local measurements are performed inside the overlap region (that is, all the four probability amplitudes $l$, $l'$, $r$, $r'$ are nonzero). The violation of the CHSH-Bell inequality identifies the presence of nonlocal correlations between the pseudospin outcomes in the regions $L$ and $R$.

Therefore, spatial overlap between wave functions associated to independently prepared identical particles can generate nonlocality effects which can be tested in quantum optical scenarios \cite{sciarrino2011bell} and then exploited to implement quantum information or communication processes\cite{LFRoperationalframe}.

\section{Conclusions}
\label{Sec:Conc}

In this work we have presented an alternative way to deal with sets of identical quantum objects. These objects, which can constitute building blocks of a complex system, can be considered as "particles" in general not coinciding with elementary excitations of quantum fields. 
This non-standard approach (NSA) to identical particles differs from the standard quantum mechanical (SA) one in that it never makes use of labels to mark particles.
The core of NSA is played by the probability amplitude between $N$-particle states. We have derived this multiparticle probability amplitude by first principles, that is \textit{cluster decomposition principle}, stating that distant experiments provide independent outcomes \cite{peresbook}, and \textit{linearity} with respect to one-particle states. The NSA is universal in that it works for any type of particle. It exhibits methodological advantages compared to the SA avoiding the problematics arising there from the necessary symmetrization with respect to unobservable labels. This occurs for example when identical particles are generated in far away regions with each particle instantaneously developing a nonzero probability amplitude of being in a space-like separated places.

The NSA also enables technical advantages with respect to SA supplying essential tools which are not usable in the latter, such as: projective measurements and partial trace. As byproducts of such tools, one straightforwardly obtains the von Neumann entropy associated to the reduced density matrix, which estimates the entanglement of a general bipartition $N$ identical particles. We have moreover introduced generalised products between vector spaces of different dimensions, showing that these can be connected with generalised annihilation and creation operators. When these operators connect states differing for one excitation, they coincides with the operators in second quantization. While in the latter the commutation rules between annihilation and creation operators derive from the field ones, here they are determined by the many-particle probability amplitude. 

We have shown that some nonidentical particle entangled states, such as the W states, cannot be used \textit{tout court} for identical particles. Within the NSA, we have introduced the spin-exchanged states (SPES) which contain the extension of W states to identical particles and allow to evidence the quantitative role of spatial overlap in the bipartite entanglement of multiparticle states. 
A new byproduct of the NSA is that indistinguishability in the presence of spatial overlap of independently prepared identical particles gives rise to exploitable nonlocal entanglement. 

Finally, the NSA provides a very convenient way for describing a system of $N$ identical particles and their entanglement. The results of this work pave the way to further studies concerning the characterization of identical particle systems, such as multipartite coherence, correlations other than entanglement and dynamics under noisy environments.

\end{document}